\documentclass{llncs}
\usepackage{graphicx}
\usepackage[italic]{mathastext}
\usepackage{adjustbox}
\usepackage{booktabs}
\usepackage{amsmath,amssymb,amsfonts}
\usepackage{subcaption}
\usepackage{setspace}

\begin{document}
\title{Achieving Observability on Fog Computing with the use of open-source tools}
%
%
\author{Breno Costa\inst{1,2} \and
Abhik Banerjee\inst{2} \and
Prem Prakash Jayaraman\inst{2} \and
Leonardo R. Carvalho\inst{1} \and
João Bachiega Jr.\inst{1} \and
Aleteia Araujo\inst{1}
}
\authorrunning{Costa et al.}
%
\institute{Department of Computer Science - University of Bras\'{i}lia (UnB) - Brasília - DF - Brazil \\ \email{\{brenogscosta,joao.bachiega.jr,leouesb\}@gmail.com and aleteia@unb.br} \and
School of Science, Computing and Engineering Technologies - Swinburne University of Technology \\ Melbourne - Victoria - Australia \email{\{abanerjee, pjayaraman\}@swin.edu.au}}
%
\maketitle              
\begin{abstract}
Fog computing can provide computational resources and low-latency communication at the network edge. But with it comes uncertainties that must be managed in order to guarantee Service Level Agreements. Service observability can help the environment better deal with uncertainties, delivering relevant and up-to-date information in a timely manner to support decision making. Observability is considered a superset of monitoring since it uses not only performance metrics, but also other instrumentation domains such as logs and traces. However, as Fog Computing is typically characterised by resource-constrained nodes and network uncertainties, increasing observability in fog can be risky due to the additional load injected into a restricted environment. There is no work in the literature that evaluated fog observability.  In this paper, we first outline the challenges of achieving observability in a Fog environment, based on which we present a formal definition of fog observability. Subsequently, a real-world Fog Computing testbed running a smart city use case is deployed, and an empirical evaluation of fog observability using open-source tools is presented. The results show that under certain conditions, it is viable to provide observability in a Fog Computing environment using open-source tools, although it is necessary to control the overhead modifying their default configuration according to the application characteristics.

\end{abstract}
%
\addtolength{\intextsep}{-6mm}
\addtolength{\abovecaptionskip}{-4mm}
\addtolength{\belowcaptionskip}{-2mm}
\addtolength{\textfloatsep}{-5mm}

\section{Introduction}
Fog computing is a computing model that brings computation and data storage closer to where it is needed, typically at the edge of the network. The main characteristics of Fog Computing are its distributed and decentralised nature, its focus on low latency and high bandwidth communication, and its ability to support a wide range of devices and applications~\cite{Bonomi2012}. Fog computing can also leverage existing network infrastructure and resources, such as routers, gateways, and edge devices, to provide a more efficient and cost-effective computing environment~\cite{iorga2018fog}.

In the context of IoT applications, Fog Computing can play a crucial role in enabling real-time data processing, analytics, and decision making at the edge of the network. This is particularly important for applications that require low-latency and high-bandwidth communication, such as industrial automation, autonomous vehicles, and smart cities~\cite{Naha2018}. In this way, Fog Computing can enable more intelligent and autonomous IoT systems, reduce network congestion and latency, and improve the overall performance and reliability of IoT applications~\cite{Naha2018}.

IoT applications can have a more fragmented code base compared to traditional applications. This is because they often involve multiple devices and systems that need to work together seamlessly, which can require different programming languages, frameworks, and protocols. In this scenario, there is increased complexity in providing system maintenance and a higher risk of not meeting service level agreements (SLAs)~\cite{vaquero2019research}.
A possible solution to deal with this scenario is to increase the observability of IoT applications to achieve timely decision-making and fast application troubleshooting \cite{pallewatta2023placement}. Observability is a concept borrowed from control theory and recently applied to the monitoring and debugging of distributed systems \cite{karumuri2021towards}. It refers to the ability to understand the internal state of a system based on its external outputs \cite{kalman1960general}. 

Monitoring is based on the collection and analysis of performance metrics from applications and helps to determine whether SLAs are guaranteed \cite{srirama2023decade}. Observability is referenced as a superset of monitoring that uses data analytics techniques on the collected monitoring data with the objective of reducing the time it takes to know why something is not working as it should \cite{marie2019demonstration}. In addition to metrics, observability also uses application logs and service call traces, allowing a cross-analysis of these data to gain more insight into system behaviour ~\cite{karumuri2021towards}, providing better opportunities for actuation, and shortening the time required to return applications to a healthy state. 
However, an observability platform must be able to manage and process large amounts of data in a timely manner, and Fog Computing is composed of resource-restricted nodes and interconnected by potentially uncertain networks. This is a critical issue that requires additional research \cite{usman2022survey}. Because of this there is a need for defining observability from a Fog Computing perspective. 

\subsection{Motivating Scenario: Smart City}
\label{subsec:Motivating}

A smart city application is software developed to help manage and improve the quality of life in a city. These applications take advantage of advanced technologies such as sensors, IoT devices, 5G networks, and artificial intelligence to collect and analyse real-time data on different aspects of the city, including public safety, traffic, energy consumption, air quality, and more~\cite{gharaibeh2017smart}. Mobile IoT-RoadBot \cite{IoT-RoadBot2022} is a Smart City application that monitors and detects roadside asset maintenance issues. Furthermore, it is used as a real-world testbed to assess 5G performance. It has been deployed on 11 waste collection service trucks (WCST) in Brimbank City Council, located within the metropolitan area of Melbourne, Australia. This testbed provided an ideal real-world deployment scenario to assess the performance of the 5G network due to the natural mobility of the trucks. It allowed the performance of a 5G network to be tested using an application that streams a great amount of real-time video data across a large geographical area. 

The 11 WCSTs have been equipped with IoT devices, including depth-sensing stereovision cameras, 5G routers, 5G dome antennas, edge computers, and Global Navigation Satellite Systems (GNSS).
Mobile IoT-RoadBot collects footage from the roadside and transmits data continuously while being in service for garbage collection around Brimbank. The application must process the video on the edge to shrink its size and aggregate context data. After this, the data are sent to the cloud to allow server-side functionalities, such as recognition of roadside assets and integration into the maintenance dashboard. WCSTs use 5G technology, but the availability of 5G networks is not ubiquitous in the region. Therefore, sometimes there will be no available connection to send data to the server, and the videos must be stored for further transmission as soon as the network becomes available. Additionally, there are other issues that can cause data flow interruption between WCSTs and servers, such as lack of space to store videos, camera malfunction, and application crash. In such cases, the delay in identifying such problems may cause data loss, money waste, and low-quality service to the citizen.

It would be beneficial for the Smart City IoT application if such issues were identified quickly and automatically. A solution could be delivered faster and the system could return to its normal state, potentially actuating autonomically~\cite{fizza2022survey}. For example, after identifying that the application is stuck, i.e. hardware resources and network are available and healthy, but the application is not collecting roadside footage nor transmitting it, it could send a remote command to restart the application and verify after a short period of time whether a different actuation is still necessary.
Thus, to provide a comprehensive and updated view of the whole system, it is necessary to increase its observability by promptly collecting the status of each component (sensors, hardware, network, application modules). These data can be processed to provide triggers for actuation, according to pre-defined thresholds and rules. Additionally, the data will be available for historical analysis.

In our previous work \cite{costa2022monitoring}, we conducted a systematic review of the literature on fog monitoring solutions. We have found that no available proposal was capable of properly managing the collection of metrics, logs, and traces simultaneously in a Fog Computing environment. This work provides a clarification of the challenges of increasing the observability of the application in this scenario. The contributions of this work are: 1. The definition of observability in a Fog Computing environment; 2. Definition of a Fog Observability Data Life Cycle and 3.
An empirical evaluation of delivering observability in a real world Fog Computing environment using a smart city use case.

This work is organised as follows.  Section \ref{sec:background} summarises the characteristics of Fog Computing and presents a definition of observability for distributed systems. The most relevant related work is provided in Section \ref{sec:Related}. Section \ref{sec:FogObservability} presents a definition of Fog Observability based on a specific data life cycle, and modelling its overhead.  A real-world Fog Computing testbed is detailed in Section \ref{sec:Experimental}. Section \ref{sec:Evaluation} describes the experiments and offers a discussion of the evaluation results. Finally, Section \ref{sec:Conclusion} concludes this work.
\\
\section{Background}
\label{sec:background}

This section presents relevant information on Fog Computing and Observability.

\label{Subsec:Fog}
\textbf{Fog Computing} - 
Fog computing was presented in 2012 \cite{Bonomi2012} with the objective of providing computing, storage and network services between end devices and cloud providers, complementing resources when it is not possible to meet the requirements with traditional cloud services. 
In recent years, the concept of Fog Computing has been improved both by academia \cite{Dastjerdi2016,Naha2018,Vaquero2014,Yi2015} and industry \cite{fognist,OpenFog17}. 
However, due to the lack of consensus on its definition in terms of scope, composition devices, architecture, service models, etc., there are some other similar paradigms, such as Edge Computing \cite{Dastjerdi2016}, Mobile Edge Computing (MEC) \cite{Dolui2017}, and Mist Computing \cite{ranaweera2021survey} that are frequently confused with fog. 
In this work we consider Fog Computing as a broader and more complete concept that can be considered as an umbrella that encompasses all other similar paradigms \cite{chiang2017clarifying}.

\begin{figure}[ht]
\centering
\includegraphics[width=0.35\textwidth]{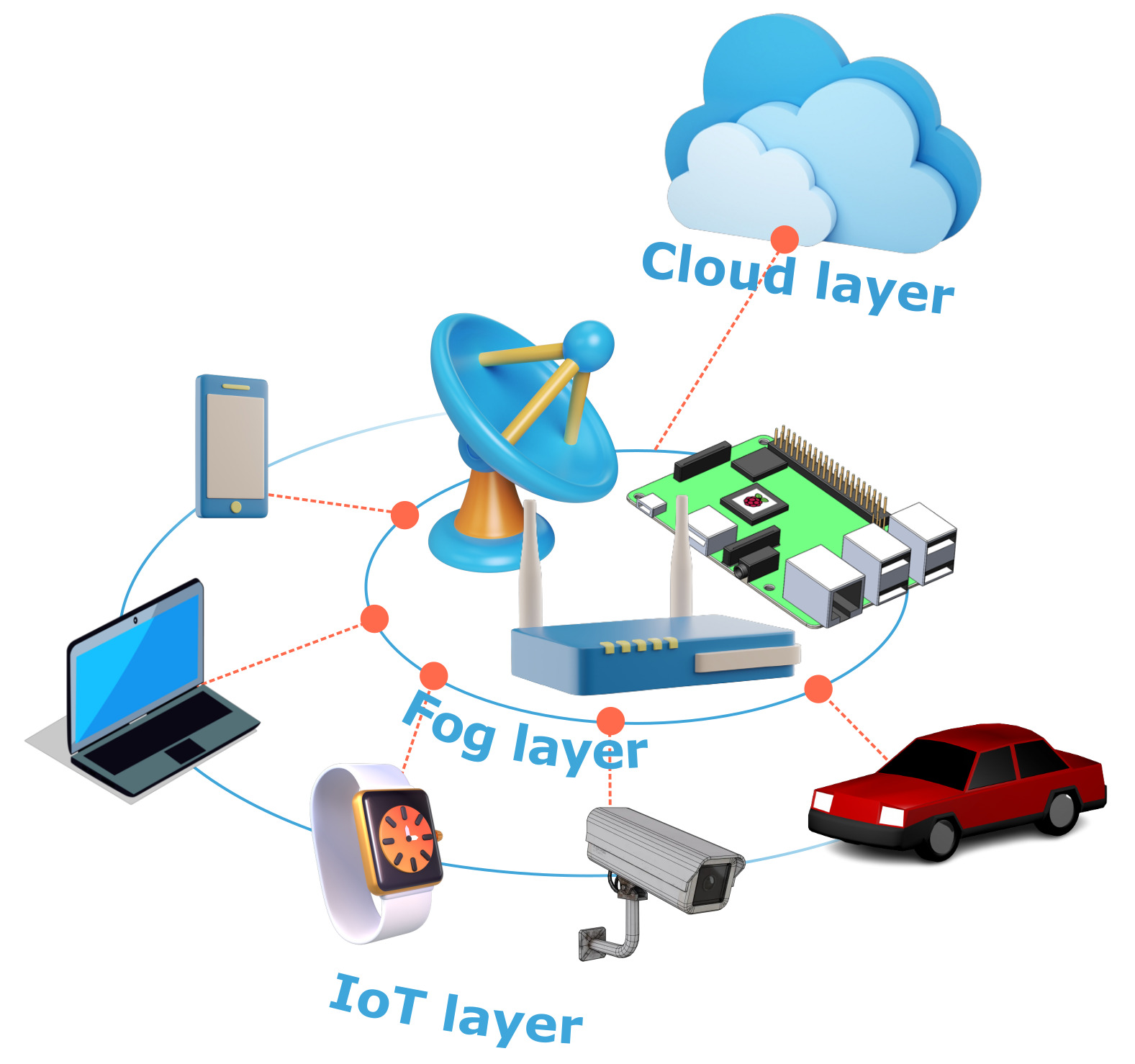}
\caption{Overview of the Fog Computing Architecture.}
\label{fig:fogoverview}
\end{figure}

The architecture most used to represent a Fog Computing environment is composed of three layers: IoT layer, Fog Layer, and Cloud Layer, as presented in Figure \ref{fig:fogoverview}.
The IoT layer represents the IoT devices connected at the edge of the network by which the end users can request the services to be processed in the above layers. The Fog Layer is placed between the IoT and Cloud Layers and provides shared resources that IoT applications can use as needed, such as processing and data storage resources, before data are transferred to the Cloud \cite{AlDoghman2016}. 
This layer is made up of nodes, commonly called \textit{fog nodes}, i.e. any hardware device that has software and hardware resources with high communication capability\cite{bachiegafognode}. 
Finally, the Cloud Layer is composed of cloud providers' services, with more robust computational resources to deliver high-order processing and long-term storage.

A Fog Computing environment 
is characterised by having a more distributed organisation, heterogeneity of physical devices and networks, and connectivity uncertainty, caused by device mobility, network instabilities, and battery exhaustion \cite{iorga2018fog}. This scenario is different from a cloud computing environment, supported by homogeneous resource-rich servers, continuous power supply, and stable redundant network connections.

\textbf{Observability} - 
Observability is a characteristic of systems that provide information about their internal states by means of external output\cite{kalman1960general}. The higher the observability, the easier it will be to understand the current and past behaviours of the system and actuate over it when needed. 
\textbf{Observability Instrumentation Domains - }
\label{subsec:domains}
Observability can be instrumented in a system by generating outputs that inform the internal state of the system at specific points in time. The different data types that compose the output are named Instrumentation Domains. Each instrumentation domain contributes to the observability of a system, offering a different perspective on the system. There is a consensus in the literature that the most important instrumentation domains of observability are metrics, logs, and traces \cite{CNCF_Whitepaper,hausenblas2021cloud,karumuri2021towards}. Some authors also consider other instrumentation domains such as events\cite{karumuri2021towards}, profiles\cite{CNCF_Whitepaper,hausenblas2021cloud} and crash dumps \cite{CNCF_Whitepaper}, although they are not recognised as such nowadays by most researchers. As time passes, some of those new domains could be standardised and incorporated into the observability's context. So we can define $ID=\{ID_1, ID_2, ID_3...ID_n\}$ as the set of instrumentation domains that are outputed by a system. Having more instrumentation domains available means a higher level of observability. Thus, observability is directly related to the cardinality of ID ($|ID|$). For instance, a Fog monitoring solution that manages only metrics has lower observability than one that manages metrics, logs, and traces simultaneously.

It is feasible to connect the instrumentation domains by the time at which each piece of information was generated. 
When it is viable to relate two or more of them in the same analysis, more opportunities for actuation arise.
In addition to the independent value of each domain, there is an additional value in the cross-analysis between domains, due to their synergistic interactions \cite{mcgrane2011method}, i.e., when two or more factors act as causes of a particular outcome. This effect is popularly known as ``The whole is more than the sum of its parts''. 
So in addition to determining the observability level of a system by the number of its instrumentation domains, we need to also consider the synergistic interactions between them as well. Synergistic interactions could be modelled as $X(ID) = ID_1~X~ID_2 ...~X~ID_n$, where $X$ is an operator that filters the data from all available instrumentation domains ($ID_i$) and returns the subset of each $ID_i$ that matches a specific period of time. Whenever more than one $ID_i$ returns a non-empty subset after applying $X$, the system has a potentially higher observability level for that period of time.
This definition shows that to increase the observability of a system it is important not only to collect information from the instrumentation domains and analyse each data set isolatedly. It is also relevant to be prepared to learn from their interactions and correlate them. 

\textbf{Metrics, Logs, and Traces} - There is a consensus in the literature that Metrics, Logs, and Traces are the most important instrumentation domains. This study will focus on them from now on.
\textbf{Metrics} are more related to the performance of a system. They are numerical values collected at a point in time and their collection can be characterised as a time series. In the motivating scenario, the following metrics are available: percentage of CPU usage,  speed of a truck in km/h, throughput of the 5G network in Mbps, amount of video data sent to the Cloud in MB, etc. 
\textbf{Logs} are unstructured or semi-structured text files that report relevant events and contextual information, and the instrumentation is usually done at the development time.
Using the motivating scenario, there is information in the logs related to the quality of service of the network connection, the geographic coordinates of the truck, etc. 
\textbf{Traces} are records of service calls made by the system. They allow observations of the call sequence delays, 
from the beginning to the end of a request. Trace analysis can show which service calls are taking longer in the response time composition of an application. They can also show requests that do not finish correctly. 
In the motivating scenario, the application performance information (upload throughput) is reported aggregated by suburb to the City Council. This aggregation took a long time to process due to the high volume (2.5 million measurements per week). After optimising the code using the point-in-polygon approach \cite{jordahl2016geopandas} instead of brute force, the time spent on this operation was reduced to 1\% of the original time.
\renewcommand{\arraystretch}{0.84}
\begin{table}[!b]
\caption{The three most important domains of observability differ in their data characteristics.}
\footnotesize
\label{tab:domains}
\centering
\begin{tabular}{@{}llll@{}}
\toprule
\textbf{Domain} & \textbf{Type}                 & \textbf{Query}             & \textbf{Storage}     \\ \midrule
Metric          & Numeric                       & Aggregations               & Time Series Database \\
Log             & Semi/not structured strings       & Approximate string search & Inverted Index       \\
Trace & DAGs of duration of execution & Disassociated graph search & Inverted Index  \\\bottomrule
\end{tabular}
\end{table}

Metrics, logs, and traces carry different types of information, as can be seen in Table \ref{tab:domains}. Each of them contributes to increasing the observability of a system, allowing for complimentary actuation.  Metrics deliver objective information about the external interface of a system, e.g., video upload throughput. 
Logs usually provide internal information about failure events, such as specific error messages, exception handling messages, and runtime errors. This information is necessary to speed up root cause analysis, help the maintenance team improve error treatment, and return the system to a healthy state. 
Traces provide details about the internal flow of information.
These data can be visualised as a graph and a critical path can be generated from it, allowing scrutiny of the dependency among the components of a distributed system \cite{sigelman2010dapper}. The volume of data depends on the amount of requests and can be bursty.

\section{Related Work}
\label{sec:Related}
This section will present relevant work on metrics, logs, and traces as isolated information domains. After that, some works concerning their integrated management in the context of observability are provided. Several works have already evaluated monitoring solutions made for cloud and prior distributed paradigms. Their results show that none of them was suitable for running in Fog environments \cite{abderrahim2017holistic,abreha2021monitoring,battula2019efficient,taherizadeh2018monitoring}. Therefore, the focus of this section will be on proposals that can be applied to Fog environments.

\textbf{Metrics, Logs, Traces} - 
Metrics, logs, and traces have been used in distributed systems for decades \cite{joyce1987monitoring} and there are several works that have studied them in isolation.
The works \cite{brogi2019measuring} and \cite{forti2021lightweight} proposed FogMon, a tool based on an agent running on the fog node, collecting metrics about the use of hardware and network resources. It relies on differential updates to reduce overall network overhead. FogMon has the purpose of monitoring performance and collects metrics at a periodic rate. The solution was evaluated in a real testbed for the additional overhead it adds \cite{gaglianese2022lightweight}. Although developed for Fog computing, the proposal does not collect logs and traces, providing a low level of observability for the systems it monitors. In \cite{casse2021using}, the authors suggest a method to connect data from multiple traces by treating all traces as a single graph and breaking down tracing information into multiple vertices and edges. The work uses container as an execution environment and has chosen the same open-source solutions for traces management, although it was evaluated in a cloud native testbed. There is no use of a data life cycle or a definition of observability.

\textbf{Observability} - 
The work \cite{karumuri2021towards} presented an industry experience of a cloud-based team collaboration service with observability data and its production use cases. The author proposed a data management architecture that decouples the real-time and historical data access tiers and the querying tier, with the aim of scaling them independently. OpenTelemetry \cite{openTelemetry} is a project that seeks to standardise the collection of observability data from cloud-native applications and their infrastructure. The goal of this project is to enable monitoring of the overall health and performance of distributed systems by gathering telemetry data.
In \cite{tzanettis2022data}, techniques to enhance the observability of distributed applications were discussed. This was done by combining multiple data sources to create a comprehensive view of the application's functioning, with the goal of improving decision-making in the orchestration process.
In \cite{levin2020viperprobe}, an extended Berkley Packet Filters (eBPF)-based microservices collection system was proposed to provide a range of metrics for single-tenant environments.
The paper \cite{usman2023desk} introduces an open-source system to control observability data of IoT applications that run on a Kubernetes edge cluster.

Unlike the described proposals, this work is focused on the challenge of achieving a higher level of observability on Fog Computing. To do so, a more formal definition of observability is proposed, as well as a Fog Observability Data Life Cycle (ODLC). Using open source tools, deployed on a fog testbed running a real-world IoT smart city application, experiments were carried out to assess how to implement an ODLC in such an environment with low overhead.
\\
\section{Fog Observability}
\label{sec:FogObservability}
Due to the Fog characteristics described in Section \ref{Subsec:Fog}, the collection, storage, and analysis of a large volume of data in Fog Computing remains an open challenge. 

\subsection{Fog Observability Data Life Cycle}
\label{subsec:cycle}
\begin{figure}[ht]
\centering
\includegraphics[width=0.75\textwidth]{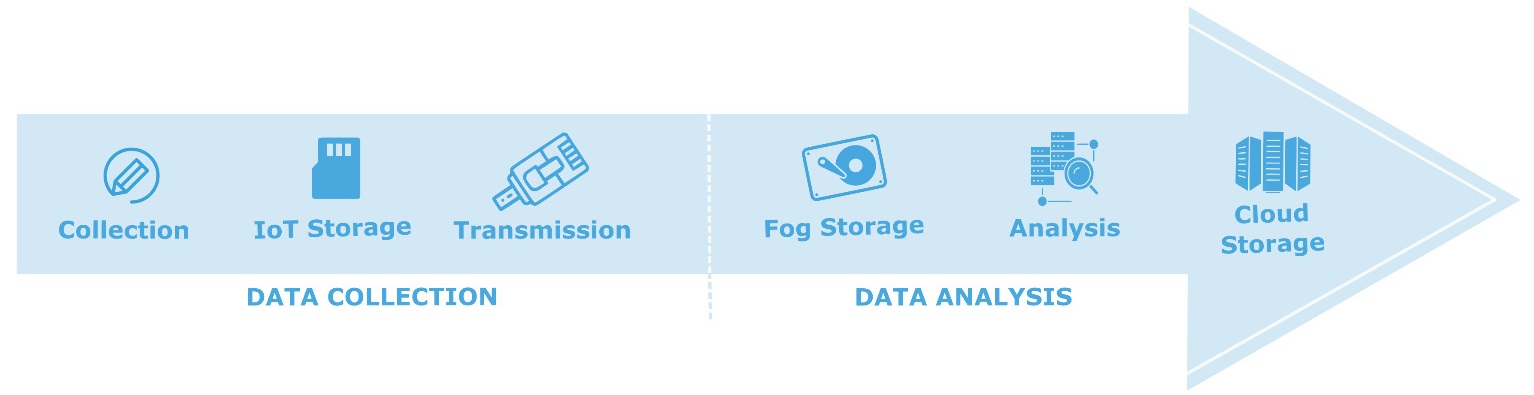}
\caption{Fog observability’s data life cycle.}
\label{Fig:LifeCycle}
\end{figure}

To obtain valuable information from each instrumentation domain and to increase the observability of an application running in a Fog environment, it is necessary to be aware of the following six-step Observability Data Life Cycle, depicted in Figure \ref{Fig:LifeCycle}:  1. Collection; 2. IoT storage; 3. Transmission of data to the Fog; 4. Fog storage; 5. Data analysis and visualisation; 6.Cloud storage and analysis. The first three Steps make up the Data Collection phase of the life cycle. The last three Steps form the Data Analysis phase.
\addtolength{\abovecaptionskip}{+3mm}
\addtolength{\belowcaptionskip}{+1mm}
\addtolength{\abovecaptionskip}{-3mm}
\addtolength{\belowcaptionskip}{-1mm}
\textbf{1. Collection} - In the initial Step of the fog observability data life cycle, the data are collected. This can happen in a multitude of ways depending on the instrumentation domain in place.  Metrics can be acquired from the operating system by means of system calls. 
Logs are written according to the specific event flow that was instrumented to be recorded in text. 
When previously instrumented, traces can be created by specific API calls that record the sequence and delay of each service call. 
 \textbf{2. IoT Storage - data staging in the device awaiting transmission} - 
Observability data are usually immutable and append-heavy \cite{karumuri2021cloud}.
In order to avoid running out of storage resources, a data removal policy should be in place.
The period of time that a device can handle stored observability data will depend on several factors, such as data footprint by period of time, the frequency of generation, and the available storage space reserved for the system. 
Although metrics can be stable in terms of data volume, logs and traces have greater variability \cite{karumuri2021towards}. 
\textbf{3. Data transmission to Fog} - Observability may allow timely and proper decision making. Although it is possible to make some minor decisions locally using a single device, critical decisions are expected to be made using a process that can assess a higher volume of data that came from different subcomponents of the system, granting a more comprehensive view of the system. 
Therefore, the data collected from the IoT layer should be transmitted to the Fog Layer, where a resource-richer node will store them and allow for a more comprehensive data analysis. 
The network connections used by the application to receive and respond to user requests may be the same as those used by the observability data flow. 
An adaptive process may be in place to define the amount of data that can be transferred from the devices, selecting which instrumentation domains will be included in each transmission, and the period of time to which the collected data will refer. 
\textbf{4. Fog Storage - Specialised pre-processing and storage according to the type of data and usage} - Fog nodes are expected to be resource richer compared to IoT devices \cite{bachiegafognode}. Due to this, it is on the Fog Layer where observability data from several IoT devices are stored with the aim of rapid actuation and decision-making. 
The metrics should be stored in a time series database (TSDB). However, logs and traces are structured differently and will benefit from other storage solutions, such as inverted index-based storage, due to the type of queries that are usually made to retrieve meaningful information from them \cite{karumuri2021towards}.
Therefore, an observability data ingestion service on the fog should consider the data requirements that each instrumentation domain needs (see Table \ref{tab:domains}), while allowing cross-analysis to be performed.
\textbf{5. Data analysis and visualisation for decision making} - Once the observability data are available on the Fog, it is possible to query them and make decisions and actuations accordingly. 
Observability data tend to give more relevant answers when they are queried as soon as they arrive, which means that most queries and analysis use more recent data (less than 24 hours) \cite{karumuri2021cloud}. Thus, it is important to guarantee fast access to this time window data. In addition to that, to save resources to continue receiving IoT data, it is important to provide automated mechanisms to send the data out of this range to long-term storage in the Cloud. 
\textbf{6. Cloud Storage - Long-term storage and historical analysis} - Cloud is the appropriate environment to store large data volumes and run heavy data processing models, such as historical analysis of observability data \cite{hashem2015rise}. 


\subsection{Fog Observability Overhead}

Observability is not an end in itself. 
The positive outcome that an increased observability level can deliver must be balanced with the amount of computational resource needed (the overhead) to reach that level. Although obvious in a generic sense, this statement gains more relevance in a Fog Computing environment due to the characteristics of resource restriction and network uncertainty. This outcome can be modelled as in Equation \ref{equ:5}:

\begin{equation}
Outcome = \frac{Observability}{Overhead}
\label{equ:5}
\end{equation}
where \textit{Overhead} is a non-zero number between 0 and 100, representing the percentage of resources (i.e. CPU, memory, network bandwidth) consumed when running the life cycle described in Section \ref{subsec:cycle}. The higher the \textit{ overhead}, the lower could be the \textit{Outcome}, considering that resource restriction is a fog characteristic.  

The observability of a system is directly related to the cardinality of ID ($|ID|$) in addition to the synergistic interactions $X(ID)$, as seen in Section \ref{sec:background}. Having $ID=\{ID_{Met}, ID_{Log}, ID_{Tra}\}$ means that the ODLC is capable of managing metrics, logs, and traces throughout the system. Due to connectivity uncertainty and resource restriction of Fog environments, part of the observability data set might be lost when there are no available resources such as storage, network bandwidth, device battery, etc. To cope with this, it is useful to define weights ($W_i$) for each instrumentation domain, representing the importance of that domain compared to the others in a given moment. As a data property, it is important to ensure that the overall sum of required weights is equal to 100\%, so $\sum{W_i} = 1$. These weights could be used to make adaptive decisions on which domain data will be collected or transmitted in the event that resources run out or in other specific situations. For example, a weight of 0\% can mean that data from that domain should not be managed. Values above 0\% will determine the rank of priorities if necessary. The weights may vary according to the needs, such as being in a regular system operation or after a massive error has occurred. 

Using the observability definition of Section \ref{sec:background} in Equation \ref{equ:5}, but replacing $ID$ by the weight set $W=\{W_{Met}, W_{Log}, W_{Tra}\}$. Splitting the resultant equation to isolate each instrumentation domain with its intrinsic overhead, and also isolating cross-analysis among the domains, the result is as follows:
\begin{equation}
Outcome = \frac{W_{Met}}{Over_{Met}} + \frac{W_{Log}}{Over_{Log}} + \frac{W_{Tra}}{Over_{Tra}} + \frac{X(ID)}{Over_{X(ID)}}
\label{equ:7}
\end{equation}
When running the observability data life cycle, the overhead of each instrumentation domain, namely, $Over_{Met}$, $Over_{Log}$ and $Over_{Tra}$ for metrics, logs, and traces, respectively, can be modelled as a function of  the amount of  resources consumed to generate, store, and transmit the information for that domain. A suggested model is to assign the maximum percentage of resource consumption among key hardware resources, such as CPU, memory, and network bandwidth, as in
$Over_{Met} = \max \{ \%CPU_{Met}, \%Mem_{Met}, \%Net_{Met} \}$ * 100. For simplicity, we consider the first 3 factors of Equation \ref{equ:7} together as representing the data collection phase. The overhead of the synergistic interactions $Over_{X(ID)}$ can be modelled in the same way, but with respect to resources consumed to store and analyse the data in the fog, and transmit them to the Cloud (Steps 4 to 6 of ODLC).

\begin{equation}
Outcome = \overbrace{
\frac{W_{Met}}{Over_{Met}} + \frac{W_{Log}}{Over_{Log}} + \frac{W_{Tra}}{Over_{Tra}}}
^{\textstyle\textnormal{DATA COLLECTION PHASE}} + 
\overbrace{\frac{X(ID)}{Over_{X(ID)}}}
^{\textstyle\textnormal{DATA ANALYSIS PHASE}}
\label{equ:8}
\end{equation}
In a dynamic Fog Computing environment, with resource restriction and network uncertainty, decision making can be more effective using a formula such as Equation \ref{equ:7} that balances the relative importance of each instrumentation domain with the effort required to collect and analyse it.

\section{Experimental Testbed}
\label{sec:Experimental}
This section presents the infrastructure used to build the testbed and carry out the experiments.

\textbf{Hardware} -
\label{subsec:Hardware}
The testbed consists of four Raspberry Pi 4 units (IoT devices), two more resourceful devices (fog nodes) and one virtual machine running in the Cloud, as
described in Table \ref{tab:hardware}.
\begin{table}[h!]
\centering
\caption{Hardware configuration for experimental testbed.}
\footnotesize
\label{tab:hardware}
\begin{tabular}{@{}llcl@{}}
\toprule
\textbf{Device name} & \textbf{Configuration} & \multicolumn{1}{l}{\textbf{Quantity}} & \textbf{Node type} \\ \midrule
Raspberry PI 4B model & Quad-core Cortex A72, 4GB RAM & 4 & IoT device \\
Fog server            & Quad-core 16GB RAM            & 2 & Fog node   \\
Virtual machine          & 8-core 32 GB RAM              & 1 & Cloud server \\ \bottomrule
\end{tabular}
\end{table}
Raspberry Pi 4 has a configuration of 4 GB of RAM and runs Ubuntu 22.04 LTS. Each Raspberry Pi 4 unit replays the data generated by a real waste collection service truck when it was in service in 2022 on the streets of Brimbank, Australia, running the Mobile-IoT-RoadBot \cite{IoT-RoadBot2022} application. These devices and the connections that link them to the Fog Layer will run the data collection phase of the ODLC described in Section \ref{subsec:cycle}. 

The next components of the testbed are the fog nodes, devices with 4 CPU and 16GB RAM, which hosts the open source tools used to store and analyse observability data. Metrics, logs, and traces come from IoT devices and are stored in the fog node. This data aggregation allows for a comprehensive view of the status of each component of this distributed smart city application. Decision making can be quick after applying simple rules on updated data, creating alerts, updating dashboards, and performing cross-analysis among domains. This is where most of the Data Analysis phase takes place.

\begin{figure}[!h]
\centering
    \begin{subfigure}[t]{0.49\textwidth}
    \centering
    \includegraphics[width=\linewidth]{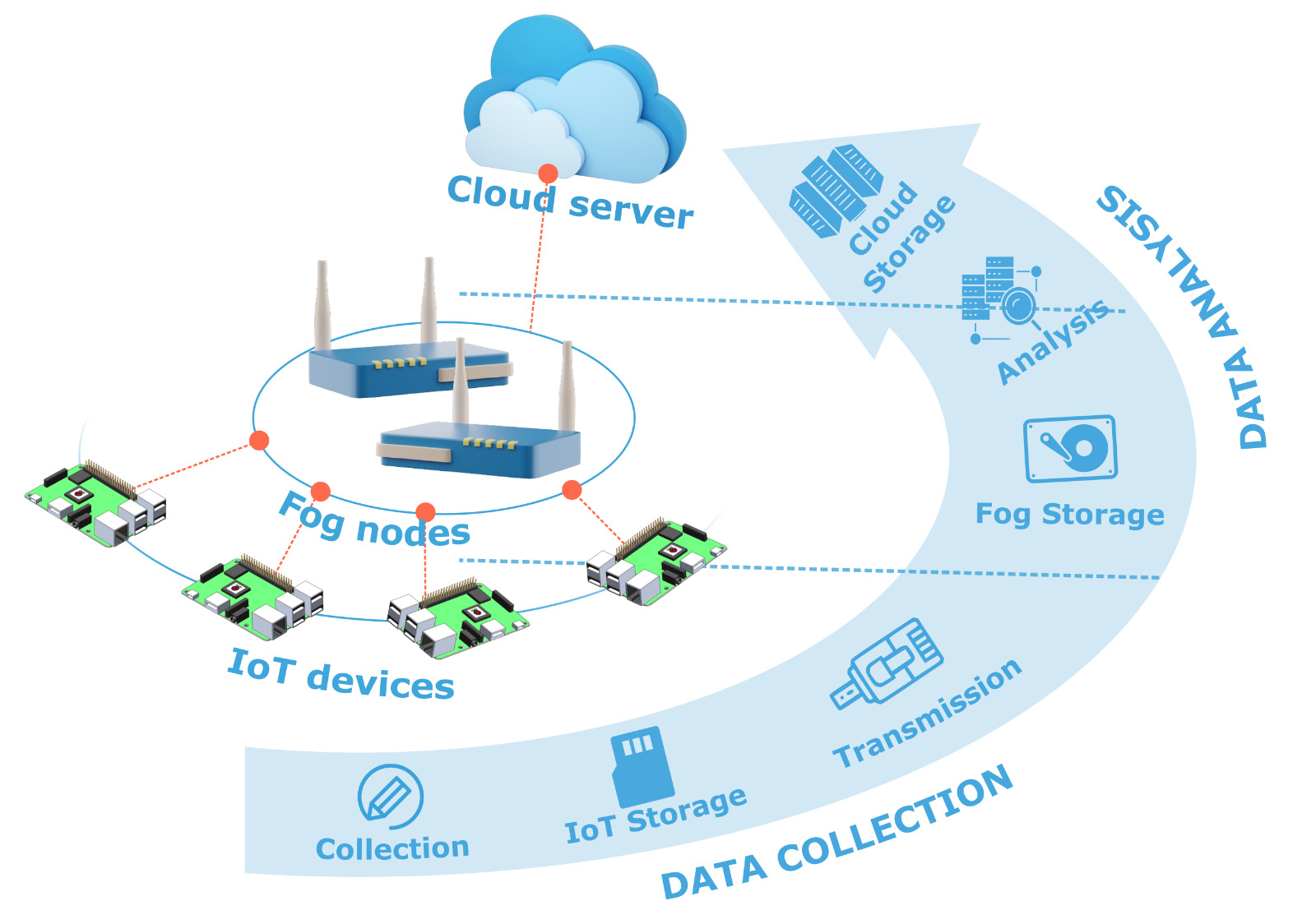}
    \caption{Testbed architecture.} 
    \label{fig:testbed}
    \end{subfigure}
\hspace{2pt}
    \begin{subfigure}[t]{0.49\textwidth}
    \centering
    \includegraphics[width=\linewidth]{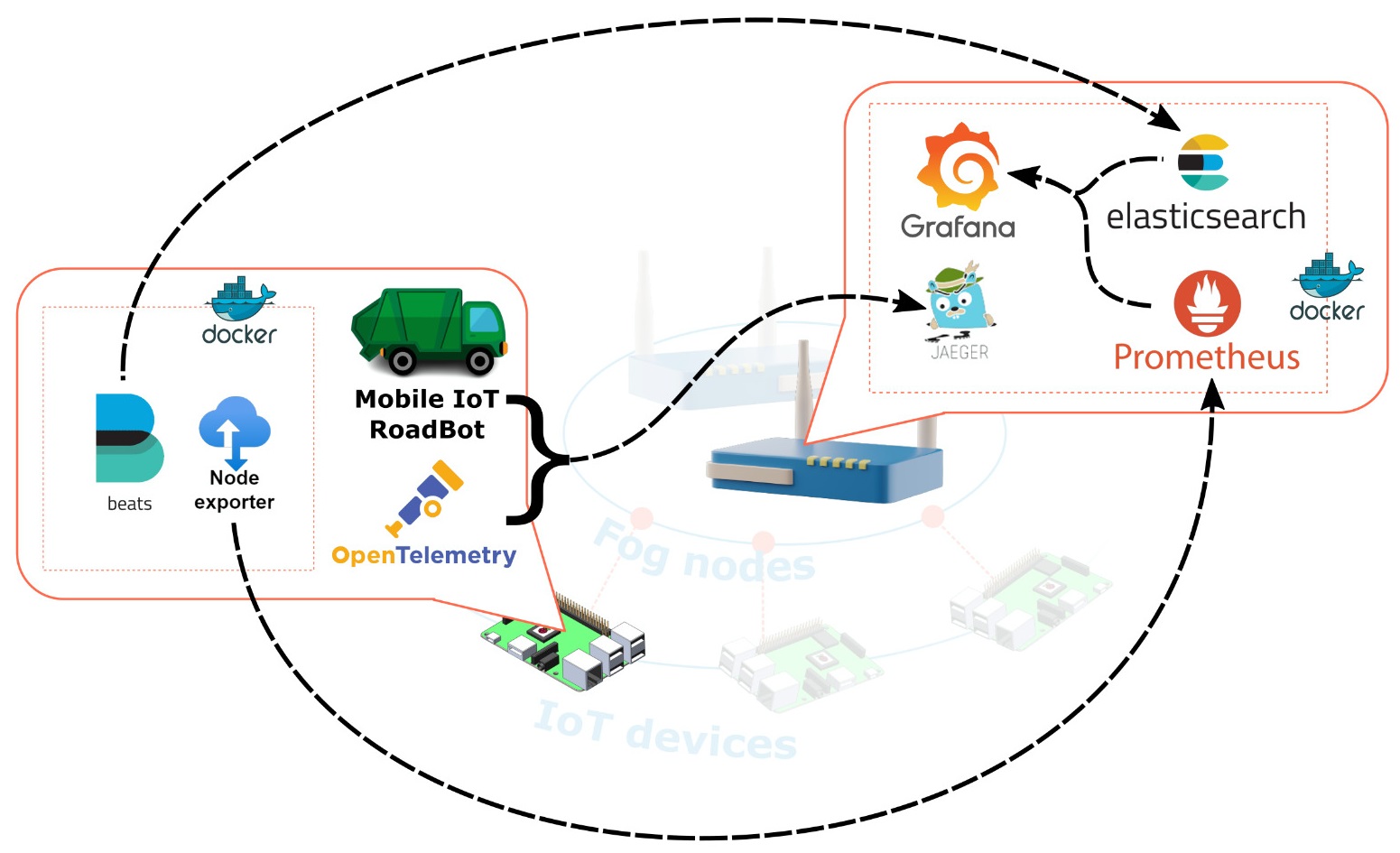}
    \caption{Observability open source tools data flow.} \label{fig:tools}
    \end{subfigure}
\caption{Fog Experimental Testbed.}
\end{figure}

The google cloud e2-standard-2 instance comprises 8 VCPUs, 32 GB RAM, and 200 GB hard disk.
The virtual machine on the Google Cloud hosts the long-term storage of observability data. Observability data are append-heavy and tend to grow indefinitely. To control the resources needed to provide fast decision-making in the Fog Layer, the volume of data should be limited. In this experiment, a week was defined as the age limit; however, this can easily be adapted to each application. Thus, every day, the data that exceed this limit are automatically moved to the Cloud. In the Cloud, a more complex and heavy analysis can be performed on historical data.
Figure \ref{fig:testbed}) shows the testbed as a 3-layer Fog Computing architecture, relating each ODLC Step to the devices where it usually occurs.
\addtolength{\abovecaptionskip}{+3mm}
\addtolength{\belowcaptionskip}{+1mm}
\addtolength{\abovecaptionskip}{-3mm}
\addtolength{\belowcaptionskip}{-1mm}

\textbf{Software} - 
\label{subsec:Software}
To our knowledge, there is no available solution that allows the collection of metrics, logs, and traces simultaneously in a Fog Computing environment \cite{costa2022monitoring}. Therefore, a solution must consist of several different tools, increasing the complexity of management. The most referenced open-source tools have been selected to characterise the challenges of increasing the observability level in a Fog Computing environment. They are described in the following paragraphs, and a summary of them is described in Table \ref{tab:software}. Every selected tool was deployed on the devices using Docker containers. The device that hosts each piece of software and its data flow can be seen in Figure \ref{fig:tools}. The virtual machine on the Cloud mimics the environment of the fog node and is not shown in the figure for simplicity. The data flow from the Fog Layer to the Cloud Layer is based on export and import operations, respectively.     
\begin{table}[]
\centering
\caption{Selected open source tools deployed on the testbed.}
\footnotesize
\label{tab:software}
\begin{tabular}{@{}llcccl@{}}
\toprule
\textbf{Tool name} &
  \textbf{Domain} &
  \textbf{\begin{tabular}[c]{@{}c@{}}IoT \\ device\end{tabular}} &
  \textbf{\begin{tabular}[c]{@{}c@{}}Fog \\ node\end{tabular}} &
  \textbf{\begin{tabular}[c]{@{}c@{}}Cloud \\ server\end{tabular}} &
  \textbf{Observation} \\ \midrule
Node Exporter     & Metrics & \checkmark &      &      & Collects metrics from the device and exposes them via an HTTP call                \\
Filebeat          & Logs    & \checkmark &      &      & Monitors specific files and transmits their content to Elastic Search.             \\
OpenTelemetry & Traces  & \checkmark &      &      & Programming language libraries to create and transmit traces to Jaeger    \\
Prometheus        & Metrics &      & \checkmark & \checkmark & Pull metrics from the Node Exporter on each IoT device \\
Elastic Search    & Logs    &      & \checkmark & \checkmark & Stores logs transmitted by Filebeat and traces data received by Jaeger            \\
Jaeger            & Traces  &      & \checkmark & \checkmark & Receives trace data transmitted by Open Telemetry SDK                             \\
Grafana           &         &      & \checkmark & \checkmark & Presents dashboards of data from Prometheus, Elastic Search and Jaeger.     \\ \bottomrule
\end{tabular}
\end{table}

\textbf{Metrics Management -} Prometheus \cite{prometheus} is the open-source standard for the collection and storage of metrics. It can be configured to pull the metrics from the IoT devices in a defined frequency (each 5s by default), and all exposed metrics are scraped and stored in a TSDB. Prometheus was deployed in the fog node. 
Grafana \cite{Grafana} is a lightweight dashboard software that is used to visualise the metrics stored in Prometheus. In this experiment, it will also connect to ElasticSearch to show log information.
Prometheus relies on external agents that run on IoT devices to generate metrics data and make them available for collection. In this work, Node Exporter is used. It is an agent that exposes hardware and network metrics such as the percentage of free CPU and the throughput of the network connection.

\textbf{Logs Management - } ElasticSearch \cite{Elastic} is an open source database of inverted index. It is used for indexing, storage, and fast retrieval of logs and traces data sent to it. Filebeat \cite{Filebeat} is an agent that runs locally on the IoT device, sending log data to ElasticSearch.

\textbf{Traces Management -} Jaeger \cite{Jaeger} was selected as an open source trace management tool. It can be configured to use ElasticSearch as storage and provides proper visualisation and fast query response time. To collect, store, and transmit traces to Jaeger, the application should be instrumented with the calls that create the traces. In this experiment, Open Telemetry SDK \cite{openTelemetry}, open source libraries available in the most used programming languages, will be used to instrument a report functionality of Mobile IoT-RoadBot.

\section{Evaluation}
\label{sec:Evaluation}
This section presents the experimental evaluation of open source observability tools using a real-world fog testbed running an IoT smart city application. The overhead of the software components and the data volume are measured. Additionally, findings on the benefits and challenges of deploying an observability framework in a Fog environment are presented and discussed. This evaluation demonstrates that an ODLC can be deployed in a Fog environment using open-source tools and shows the challenges and design decisions that should be in perspective.

\subsection{Methodology}
\label{subsec:methodoloy}
To assess the overhead of achieving a higher level of observability in a Fog environment, the testbed was used to measure: 1. The overhead of each tool in the IoT devices; 2. The overhead of the server component of each tool running on a fog node; 3. Volume of observability data collected, stored, and transmitted through the data life cycle. 
While Mobile IoT-RoadBot was running on a Raspberry Pi, we deployed each agent on the same device, and measured the overhead in terms of CPU and memory usage for 4 hours. We use the System Activity Reporter (SAR) \cite{Sar} to get the average CPU and memory usage each 5 minutes. After recording this information, we uninstalled the tool and repeated the process with the other tools (Node Exporter, Filebeat, and OpenTelemetry SDK, described in Section \ref{subsec:Software}).
We performed a similar procedure related to the server side of the tools (Prometheus, ElasticSearch, and Jaeger, described in Section \ref{subsec:Software}). 
After measuring the overhead, we run Mobile IoT-RoadBot for 9 hours (a regular whole-day journey for each truck) to assess the volume of data collected by each instrumentation domain and transmitted through the layers of the testbed architecture.
Finally, to assess the benefits of achieving a higher level of observability on Fog environments, we replayed by 9 hours on the testbed the real-world output of 4 WCSTs. Each real-world WCST output is replayed by a Raspberry Pi that also runs observability agents sending data through ODLC from the IoT Devices to the Cloud server.
\subsection{Overhead}
\label{subsec:overhead}
\begin{figure}[t!]%
\centering
    \begin{subfigure}[t]{0.26\textwidth}
        \centering
        \includegraphics[width=\linewidth] {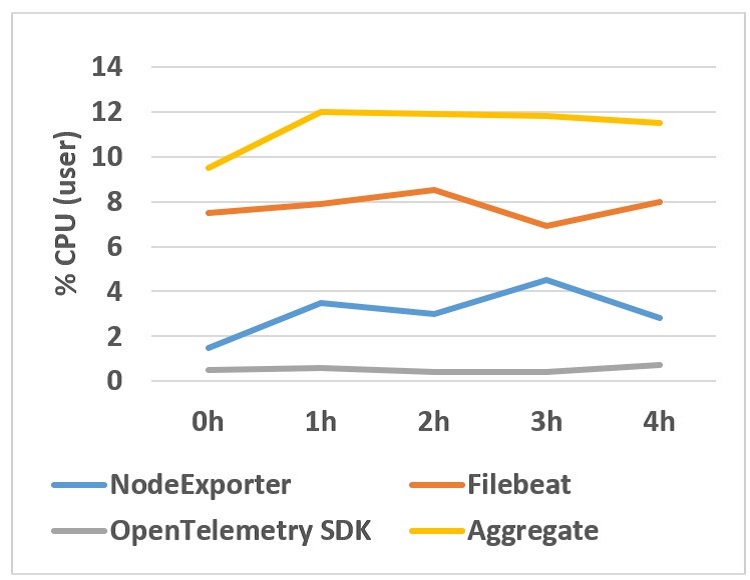}%
       \caption{CPU overhead on IoT.}%
        \label{fig:iot_cpu_over}%
     \end{subfigure}%
     \hspace{1pt}
     \begin{subfigure}[t]{0.26\textwidth}
         \centering
         \includegraphics[width=\linewidth]{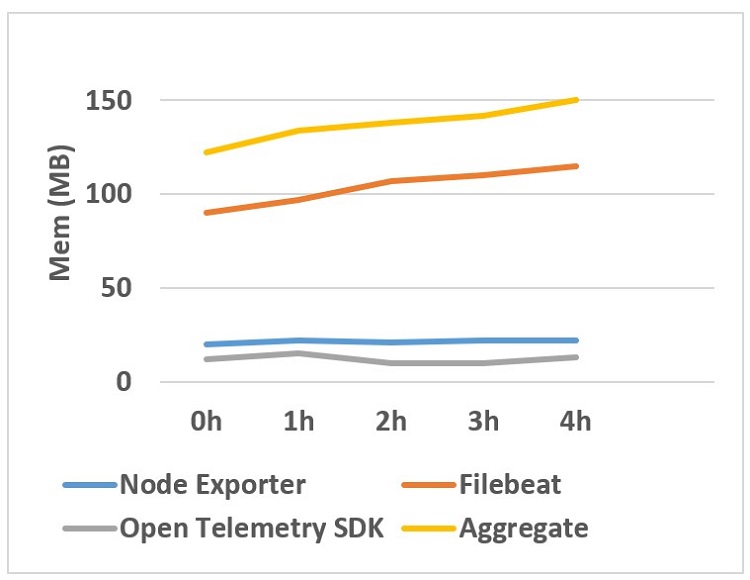}%
         \caption{Memory overhead on IoT.}%
         \label{fig:iot_mem_over}%
     \end{subfigure}%
\\
     \hspace{1pt}
\begin{subfigure}[t]{0.26\textwidth}
         \centering
         \includegraphics[width=\linewidth]{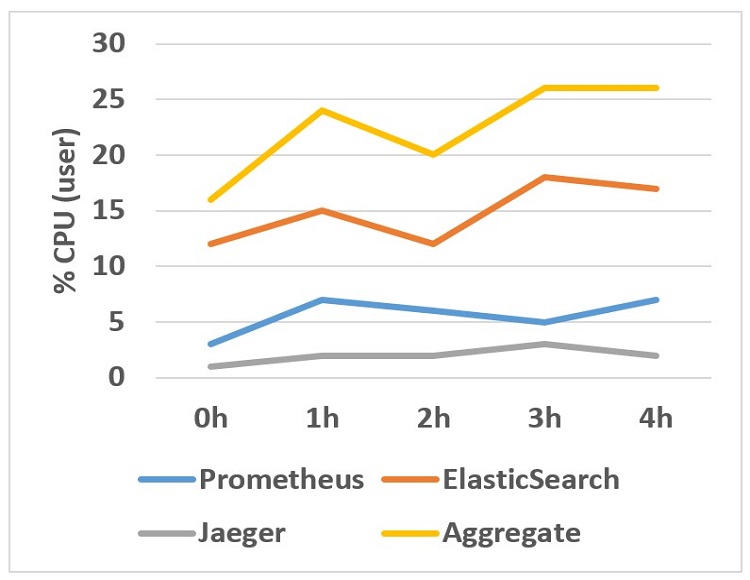}%
         \caption{CPU overhead on fog.}%
         \label{fig:fog_cpu_over}%
     \end{subfigure}%
     \hspace{1pt}
    \begin{subfigure}[t]{0.26\textwidth}
         \centering
         \includegraphics[width=\linewidth]{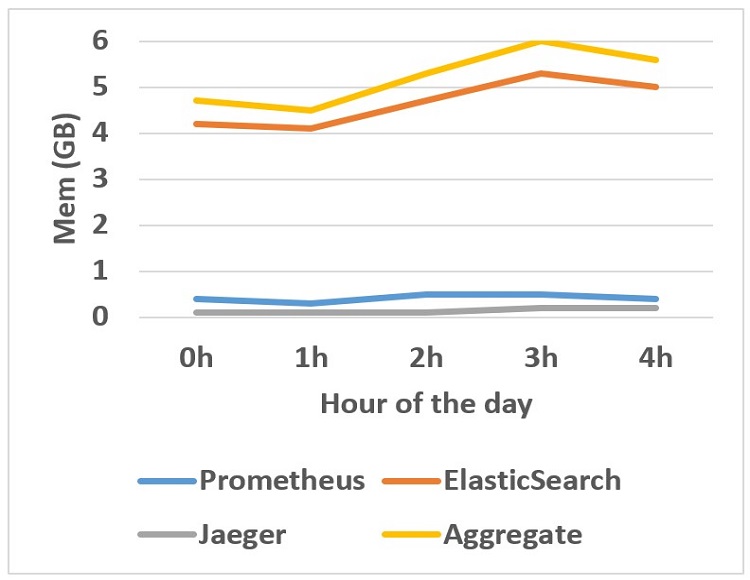}%
         \caption{Memory overhead on fog.}%
         \label{fig:fog_mem_over}%
     \end{subfigure}%
     \caption{Assessing overhead of observability tools on: IoT devices (a) and (b); Fog Nodes (c) and (d).}
        \label{fig:overhead}
\end{figure}

Figure \ref{fig:overhead} shows the CPU and memory usage of each component of the open source observability tool set being evaluated in this work. Figures \ref{fig:iot_cpu_over} and \ref{fig:iot_mem_over} show the overhead of the IoT devices. We observe negligible CPU and memory overhead. The aggregated amount of CPU when all three components, i.e. NodeExporter, Filebeat, and OpenTelemetry SDK, run simultaneously is under 12\% on average. In the case of memory usage, an aggregated footprint of less than 150MiB of RAM is required.
\addtolength{\abovecaptionskip}{+3mm}
\addtolength{\belowcaptionskip}{+1mm}
\addtolength{\abovecaptionskip}{-3mm}
\addtolength{\belowcaptionskip}{-1mm}
Figures \ref{fig:fog_cpu_over} and \ref{fig:fog_mem_over} show the overhead on the Fog node. Unlike the IoT devices, CPU usage is higher. This is expected since the Fog node deals with 4 times the volume of data (sent by the 4 IoT devices), with the purpose of receiving, processing and storage of data. A CPU usage average of less than 25\% for this kind of processing seems to be worth it. However, it prevents the server side of the observability tool set from being on the IoT layer, where devices have less resources. To guarantee steady performance, the volume of stored data was limited to a week.
In terms of memory usage on fog nodes, Prometheus allocated 400 MiB of RAM on average, while Jaeger allocated around 200MiB, a very low memory footprint for the load of Mobile IoT-RoadBot observability data. However, ElasticSearch allocated almost 4.5GB of RAM. Mobile IoT-RoadBot has a steady data collection flow and usually does not generate peaks of transmitted data. However, when dealing with a more data-intensive application or an application that has a bursty behaviour, the server side overhead should be monitored to guarantee that it copes with the needed load.
\subsection{Cross Analysis of Metrics, Logs and Traces on the Fog}
In the last subsection, the overhead added to the Fog infrastructure after deploying open source observability tools that implement an ODLC (Section \ref{subsec:cycle}) was detailed. 
This subsection will show the benefits that Mobile IoT-RoadBot could have if it was using such a set of observability tools.
\addtolength{\abovecaptionskip}{+3mm}
\addtolength{\belowcaptionskip}{+1mm}
\begin{figure}[h!]%
\centering
    \begin{subfigure}[t]{0.36\textwidth}
        \centering
        \includegraphics[width=\linewidth] {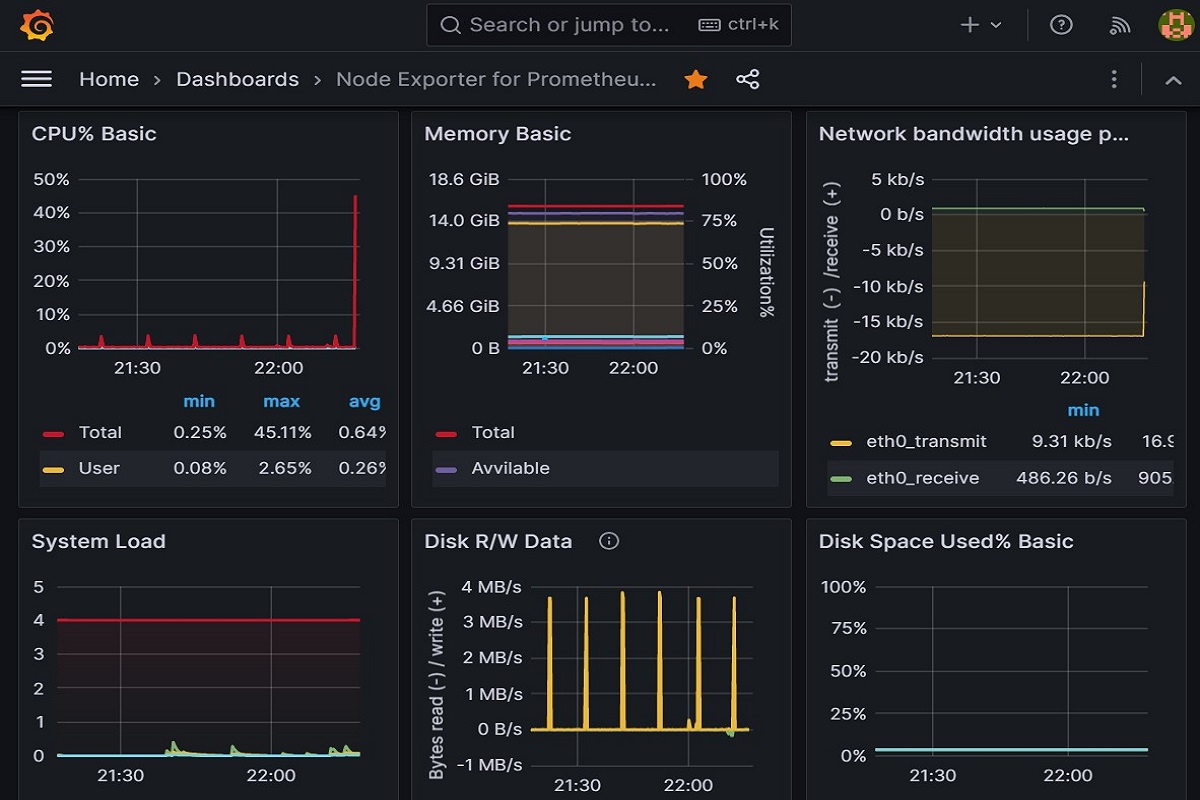}%
       \caption{}%
        \label{fig:met_fog}%
     \end{subfigure}%
     \hspace{1pt}
     \begin{subfigure}[t]{0.36\textwidth}
         \centering
         \includegraphics[width=\linewidth]{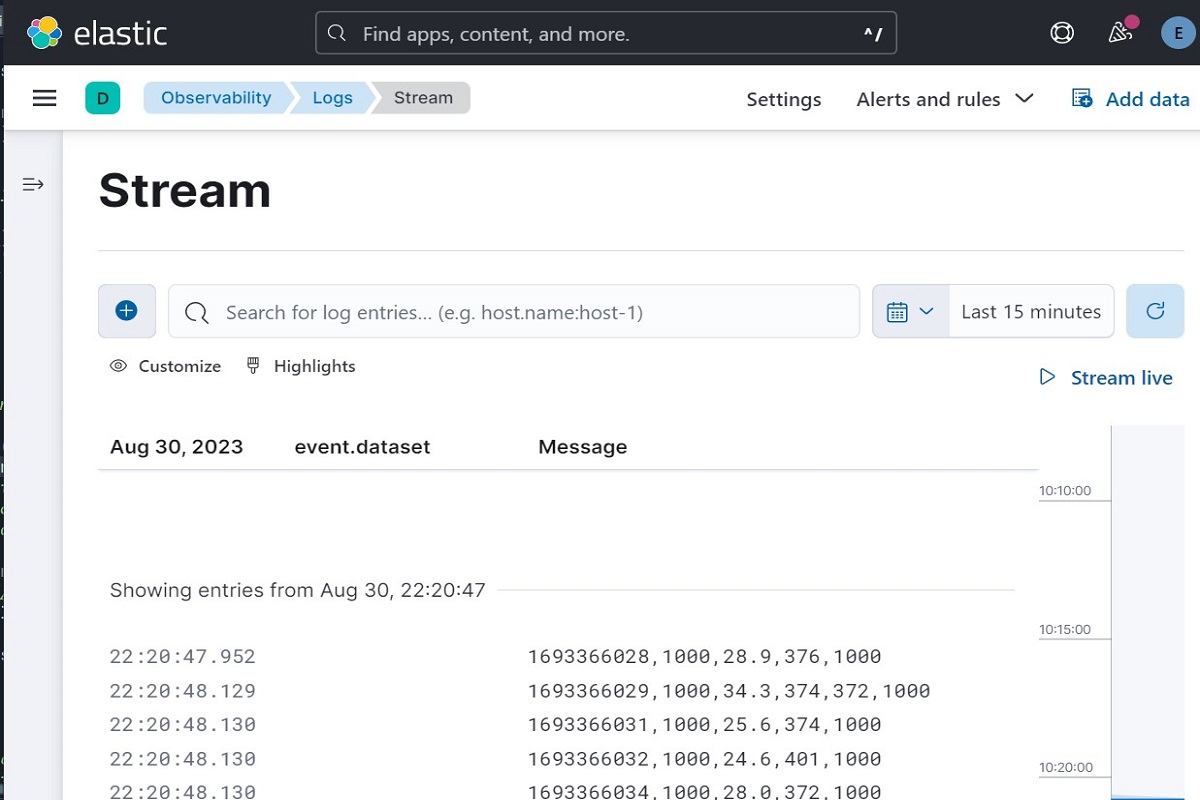}%
         \caption{}%
         \label{fig:logs_fog}%
     \end{subfigure}%
     \hspace{1pt}
\begin{subfigure}[t]{0.36\textwidth}
         \centering
         \includegraphics[width=\linewidth]{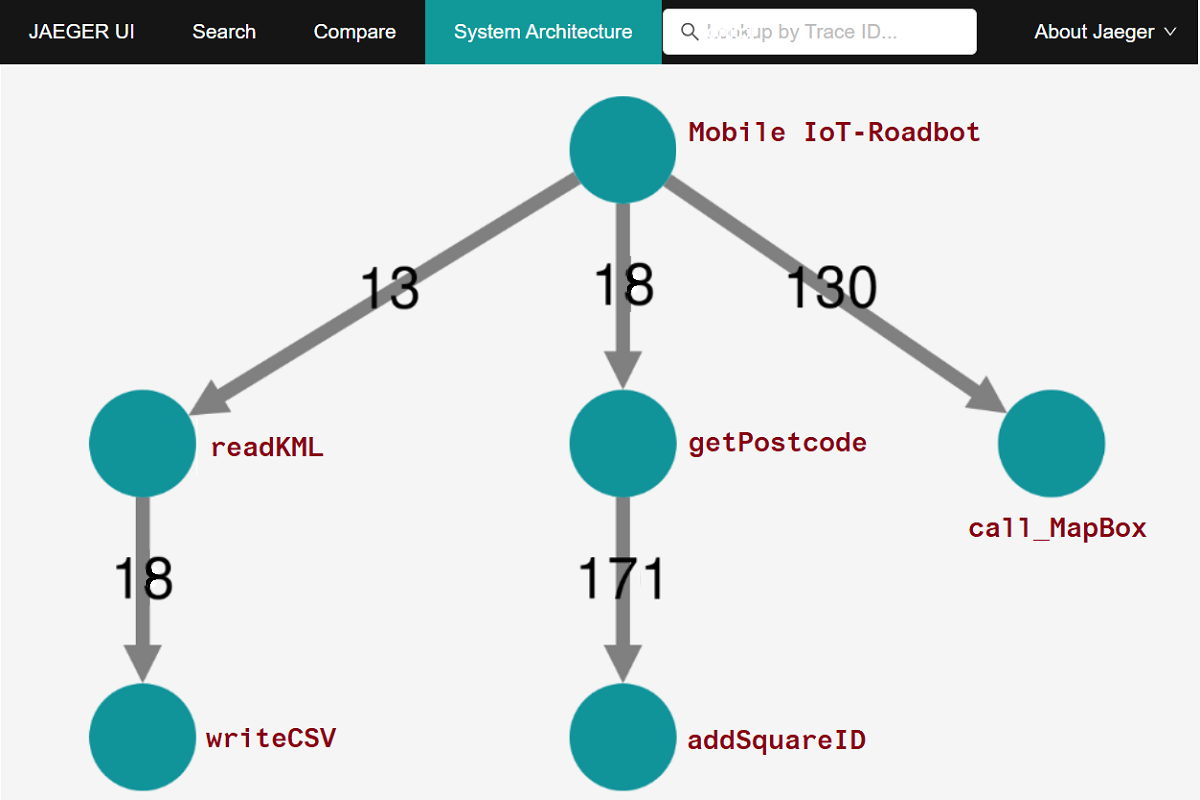}%
         \caption{}%
         \label{fig:traces_fog}%
     \end{subfigure}%
     \hspace{2pt}
    \begin{subfigure}[t]{0.36\textwidth}
         \centering
         \includegraphics[width=\linewidth]{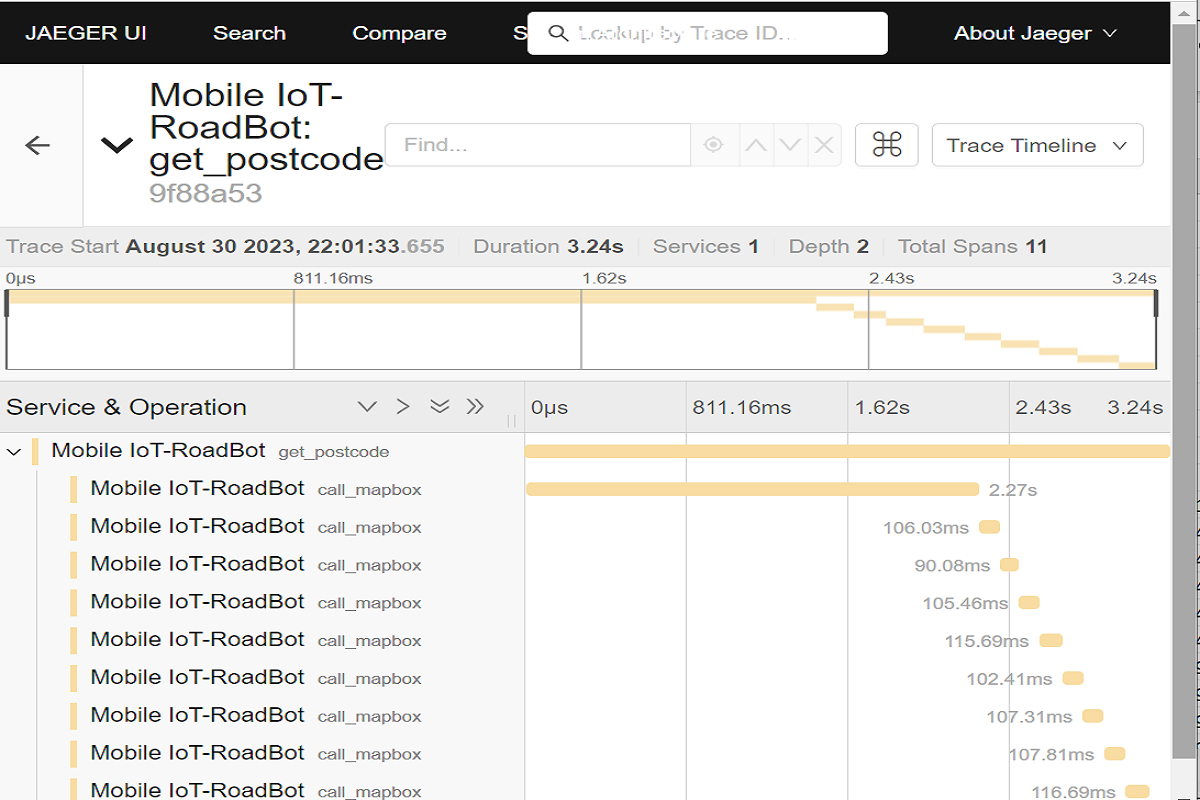}%
         \caption{}%
         \label{fig:traces_fog2}%
     \end{subfigure}%
     \caption{Observability data visualisation on Fog Layer: Metrics (a), Logs (b), and Traces (c) and (d).}
        \label{fig:visualisation}
\end{figure}
\addtolength{\abovecaptionskip}{-3mm}
\addtolength{\belowcaptionskip}{-1mm}
Figure \ref{fig:visualisation} shows how the observability data can be analysed on the Fog using the open source framework deployed on the testbed. Figure \ref{fig:met_fog} shows the value of some metrics (CPU, memory and bandwidth usage) collected by the experiments in the last hour from IoT devices. If any metric is outside the range considered safe, an alert message could be sent to the maintenance crew, allowing quick action. Figure \ref{fig:logs_fog} shows the visualisation of logs collected from IoT devices each 1 second. The logs can be easily preprocessed while being ingested to identify different fields of information, providing fast queries and allowing for alert management. Figure \ref{fig:logs_fog} shows the result of a query made on log data, detailing the latency between the IoT device and some servers of interest on the Internet.  Figure \ref{fig:traces_fog} shows a dependency graph, created from the collected traces, where one can see the average delay of each dependent component after hundreds of requests and identify which cause the majority of the response time. This information is relevant for planning future performance improvements. Finally, Figure \ref{fig:traces_fog2} shows the details of a specific trace, where it is possible to identify the components that cause the longest response time. When looking for the root cause of an identified issue, this information is very useful. 
\subsection{Discussion}
\label{subsec:discussion}
\begin{figure}[!h]
\centering
\adjustbox{trim={.00\width} {.0\height} {0.0\width} {.5\height},clip}%
  {
\includegraphics[width=0.70\textwidth]{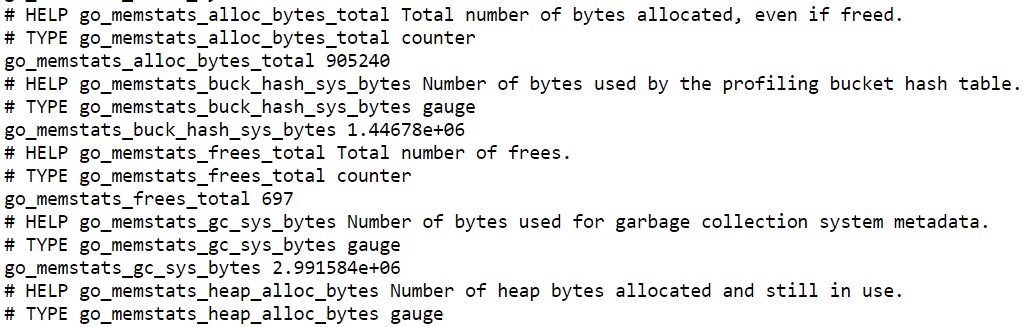}}
\caption{Snippet of default output of NodeExporter.} \label{fig:help_text}
\end{figure}

Table \ref{tab:data} presents the volume of observability data that were managed by the data life cycle during the experiments.
NodeExporter was deployed with the default configuration. Although it is a tool with a small footprint in terms of CPU and memory usage \cite{gro2018comparison}, it may have a not negligible impact in terms of the volume of data it collects. The default set of metrics that it exposes accounts for 65KB of information. These data are presented only when Prometheus pulls them using an HTTP call. This means that there is no IoT storage for these data. Prometheus is configured by default to scrape the NodeExporter page every 5s, getting all the metrics exposed and storing them in its TSDB on the Fog node. Considering that there are four IoT devices exposing metrics, the data volume transmitted to and stored on the Fog node is about 8.75GB in the span of a week, the period when these data will be available for decision-making and other analysis on the Fog Layer. After reaching a week of age, the information is removed from the fog node and sent to the Cloud for long-term storage and historical analysis. As a matter of estimation, the volume on the Cloud will reach 75GB after 2 months of operation.
\begin{table}[b]
\centering
\caption{Mobile IoT-Roadbot assessment of each observability domain .}
\footnotesize
\label{tab:data}
\begin{tabular}{@{}llllllllll@{}}
\toprule
\textbf{Tool} &
\textbf{Domain} &
\begin{tabular}[c]{@{}l@{}}\textbf{Data}\\ \textbf{Collection}\end{tabular} &
  \textbf{Frequency} &
  \begin{tabular}[c]{@{}l@{}} \textbf{Volume}\\\textbf{by Hour} 
  \end{tabular} &
  \begin{tabular}[c]{@{}l@{}}\textbf{IoT}\\\textbf{Storage}
  \end{tabular} &
  \begin{tabular}[c]{@{}l@{}}\textbf{Fog}\\  \textbf{Storage}\end{tabular} &
  \begin{tabular}[c]{@{}l@{}}\textbf{Fog Volume}\\\textbf{(1 week)}\end{tabular} &
  \begin{tabular}[c]{@{}l@{}}\textbf{Cloud}\\ \textbf{Storage}\end{tabular} &
  \begin{tabular}[c]{@{}l@{}}\textbf{Cloud Vol.}\\ \textbf{(2 months)}\end{tabular}\\
  \midrule
\begin{tabular}[c]{@{}l@{}}Node Exporter\end{tabular} &
  Metrics &
  65KB &
  each 5s &
  46 MB &
  No &
  Yes &
  8.75 GB &
  Yes &
  75 GB \\
Filebeat &
  Logs &
  1KB &
  each 1s &
  3.50 MB &
  Yes &
  Yes &
  0.67 GB &
  Yes &
  5.77 GB \\
{\begin{tabular}[c]{@{}l@{}}Open Telemetry \end{tabular}} &
  Traces &
  4KB &
  each 15s &
  1 MB &
  No &
  Yes &
  0.2 GB &
  Yes &
  1.54 GB \\ \bottomrule
\end{tabular}
\end{table}
The default output from NodeExporter provides help text for each metric, as shown in Figure \ref{fig:help_text}. This information accounts for at least 20\% of the total output footprint and should be removed prior to exposing the metrics. The default set of metrics is very extensive and probably not all metrics are useful for every use case. For example, the node exporter exposes dozens of Go environment metrics (Figure \ref{fig:help_text}) that are not of interest for the monitoring of Mobile IoT- RoadBot and should be removed.  In addition to cutting off metrics that are not of interest, machine learning over historical data can be used to figure out metric correlations and keep the target metric set at minimum \cite{bali2019rule}. Furthermore, the frequency of scraping can be decreased in the Prometheus configuration without a relevant loss of opportunities for actuation. Increasing the scrap delay to 10 seconds will reduce the data volume transmitted to and stored on the Fog Layer by half. 
Using the strategies of removing the help text and changing the configuration of Node Exporter to expose only metrics about CPU, memory, disk, network, and power supply, and increasing the scrap delay to 10 seconds on Prometheus, we could reduce the volume of metric data on the Fog node by 87\%, which also positively affected CPU and memory usage by Prometheus.
Regarding the logs generated by Mobile IoT-Roadbot while the trucks were moving around the city, they record information about 5G network analysis, such as latency and throughput, and contextual information (GNSS coordinates, truck speed, etc.). Although the application writes information in the logs every second, the volume of data written is low (0.67 GB, in the span of a week, as seen in Table \ref{tab:data}), being smaller than the volume of data generated by Node Exporter after applying volume reduction strategies. Filebeat was configured to harvest only the logs written by Mobile IoT-RoadBot and transmit them to the Fog Layer. It was necessary to change the default configuration of Filebeat to turn off the auto-discovery feature. When this feature is active, Filebeat receives from the Docker manager every status change of any container on the device, consuming more memory than necessary.  As the application was not originally instrumented to record trace calls, we instrumented it in a reporting feature, utilised to aggregate 5G data by the suburbs of Brimbank. To make this aggregation, geographic coordinates were used to find the full address of Australia using a service called MapBox \cite{Mapbox}. Using the data footprint of these traces, we estimate the data volume to generate the traces of regular operation of the Mobile IoT-RoadBot. This use case does not have bursty behaviour in terms of request processing because it performs the same volume of operations while in service. Therefore, the volume of trace data is steady.

The aggregated data volume, collected by the four IoT devices, transmitted and stored on the fog node for a period of one week was approximately 10GB, considering the default configuration of the open source tools used as shown in Table \ref{tab:data}. Using the strategies described above, the volume of aggregated data was reduced to 2GB, a reduction of 80\%. The four trucks whose observability data are replayed by the IoT devices in this experiment transmitted 291 GB of video data using the 5G network in a week of real world operation. Therefore, the observability data (2GB) would represent an overhead of less than 1\% in this use case. 
The experiments show that it is possible to collect the benefits of achieving a higher level of observability for a system in a Fog computing environment. In addition, the overhead of deploying an observability data life cycle can be low, if properly managed. The utilisation of Docker containers as the runtime environment for the observability tools help to address the Fog challenge of device heterogeneity. Due to the resource restriction of IoT devices, observability data collection should be done by lightweight agents. In addition, the data footprint should be minimised to reduce the risk of network congestion and increased overhead to collect and transmit the data to the Fog Layer. Each instrumentation domain has specific data requirements (Table \ref{tab:domains}) that must be met to optimise storage and minimise the average delay in analysing the observability data in the Fog Layer for decision-making and actuation. Leaving in the Fog Layer only a window of most recent observability data is another strategy to cope with the resource-restriction of fog nodes. Data that are outside the age range are sent to the Cloud for long-term storage and historical analysis.  

The open source tools selected to make up the experimental setup are managed independently. This scenario makes more complex actuation difficult to implement. For instance, in the dynamic environment of Fog Computing, a system may present errors running on specific devices while it is functioning properly on others. In such cases, if there are not enough resources to transmit all observability data to the Fog Layer, a proper decision should be prioritising data from those specific devices and returning to regular operation when the issue is solved. To implement such adaptive and autonomous behaviour, it might be necessary to orchestrate the observability data life cycle and its agents. To our knowledge, there is no Fog solution in the literature that provides this functionality \cite{costa2022monitoring}. 
\\\section{Conclusion and Future Work}
\label{sec:Conclusion}
This work defined the characteristics and challenges of increasing the observability of systems in a Fog Computing environment. The higher the observability, the more likely it is to understand the root cause of issues in the runtime environment and solve those issues more quickly, helping to guarantee that SLAs are met. A definition of observability aligned with the literature and considering Fog challenges is proposed. 
Observability can be instrumented using metrics, logs, and traces. Each of these observability instrumentation domains has specific characteristics with respect to the type of data, the volume, and the frequency of collection. Specific mechanisms are needed to collect this information simultaneously and store it centrally, implementing a data life cycle that allows rapid decision-making, even autonomously, and that meets the requirements imposed by the Fog environment.
An IoT Smart City use case was deployed in a real Fog Computing testbed, configured with open source observability tools, to characterise the overhead that these tools add to the infrastructure and to understand the challenges that must be faced to increase the observability of a system in this scenario. The results show that although none of the selected tools was developed specifically for Fog computing, it was possible to collect the benefits of increased observability with a minimum overhead, after using strategies to reduce data volume while maintaining opportunities for actuation.
Future work may propose appropriate support to overcome the identified challenges. 
\\\section{Acknowledgements}
\label{sec:Ack}
This study was financed in part by the Coordenação de Aperfeiçoamento de Pessoal de Nível Superior - Brasil (CAPES) - Finance Code 001, in the context of the Capes PrInt programme. This work is based upon work supported by the Google Cloud Research Credits programme with the award ``sabreenar 216815189''. \\
\begin{spacing}{0.9}
\bibliographystyle{splncs04}
\bibliography{article}
\end{spacing}

\end{document}